\documentclass[preprint2,numberedappendix]{emulateapj-rtx4}
\usepackage{graphicx,natbib,bm,url,color}
\graphicspath{{./fig/}{./png/}}

\newcommand{\EQ}{\begin{equation}}
\newcommand{\EN}{\end{equation}}
\newcommand{\EQA}{\begin{eqnarray}}
\newcommand{\ENA}{\end{eqnarray}}

\newcommand{\EEq}[1]{Equation~(\ref{#1})}
\newcommand{\Eq}[1]{Equation~(\ref{#1})}
\newcommand{\Eqs}[2]{Equations~(\ref{#1}) and~(\ref{#2})}

\newcommand{\eqs}[2]{(\ref{#1}) and~(\ref{#2})}

\newcommand{\Sec}[1]{Section~\ref{#1}}

\newcommand{\Fig}[1]{Figure~\ref{#1}}

\newcommand{\Figp}[2]{Figure~\ref{#1}({#2})}
\newcommand{\Figsp}[3]{Figures~\ref{#1}({#2}) and ({#3})}
\newcommand{\Figssp}[3]{Figures~\ref{#1}({#2})--({#3})}
\newcommand{\Tab}[1]{Table~\ref{#1}}

\newcommand{\bra}[1]{\langle #1\rangle}

\newcommand{\meanrho}{\overline{\rho}}

{}
{}
{}
{}
{}

{}
{}
{}
{}
{}
{}
{}
{}
{}
{}
{}
{}
{}
{}
{}
{}
{}
{}
{}
{}
{}

{}
{}
{}

{}

{}
{}

\newcommand{\kk}{\bm{k}}

\newcommand{\xx}{\bm{x}}

\newcommand{\BB}{\bm{B}}

\newcommand{\uu}{\mbox{\boldmath $u$} {}}
\newcommand{\UU}{\mbox{\boldmath $U$} {}}

\newcommand{\AAA}{\mbox{\boldmath $A$} {}}

\newcommand{\nab}{\mbox{\boldmath $\nabla$} {}}

\newcommand{\SSSS}{\mbox{\boldmath ${\sf S}$} {}}

\newcommand{\ii}{{\rm i}}

\newcommand{\DD}{{\rm D} {}}

\newcommand{\dd}{{\rm d} {}}

\newcommand{\const}{{\rm const}  {}}

\def\la{\mathrel{\mathchoice {\vcenter{\offinterlineskip\halign{\hfil
$\displaystyle##$\hfil\cr<\cr\sim\cr}}}
{\vcenter{\offinterlineskip\halign{\hfil$\textstyle##$\hfil\cr<\cr\sim\cr}}}
{\vcenter{\offinterlineskip\halign{\hfil$\scriptstyle##$\hfil\cr<\cr\sim\cr}}}
{\vcenter{\offinterlineskip\halign{\hfil$\scriptscriptstyle##$\hfil\cr<\cr\sim\cr}}}}}

\def\Rey{\mbox{\rm Re}}

\def\Lu{\mbox{\rm Lu}}

\def\cs{c_{\rm s}}
\def\xiM{\xi_{\rm M}}

\def\vA{v_{\rm A}}

\def\kB{k_{\rm B}}

\def\kone{k_1}
\def\vsat{v_\lambda}

\def\urms{u_{\rm rms}}

\def\half{{\textstyle{1\over2}}}

\def\onethird{{\textstyle{1\over3}}}

\newcommand{\G}{\,{\rm G}}

\newcommand{\K}{\,{\rm K}}
\newcommand{\g}{\,{\rm g}}
\newcommand{\s}{\,{\rm s}}

\newcommand{\cm}{\,{\rm cm}}

\newcommand{\GeV}{\,{\rm GeV}}

\newcommand{\Mpc}{\,{\rm Mpc}}

\newcommand{\erg}{\,{\rm erg}}

\newcommand{\alphaem}{\ensuremath{\alpha_{\rm em}}}

\newcommand{\yapj}[3]{ #1, {ApJ,} {#2}, #3}

\newcommand{\yapjl}[3]{ #1, {ApJ,} {#2}, #3}

\newcommand{\yana}[3]{ #1, {A\&A,} {#2}, #3}

\newcommand{\yanar}[3]{ #1, {A\&A Rev.,} {#2}, #3}

\newcommand{\yprl}[3]{ #1, {Phys.\ Rev.\ Lett.,} {#2}, #3}

\newcommand{\ymn}[3]{ #1, {MNRAS,} {#2}, #3}
\newcommand{\ynat}[3]{ #1, {Nature,} {#2}, #3}

\newcommand{\ypr}[3]{ #1, {Phys.\ Rev.,} {#2}, #3}

\newcommand{\yprd}[3]{ #1, {Phys.\ Rev.\ D,} {#2}, #3}
\newcommand{\ypre}[3]{ #1, {Phys.\ Rev.\ E,} {#2}, #3}

\newcommand{\yjhep}[3]{ #1, {JHEP,} {#2}, #3}

\newcommand{\yjour}[4]{ #1, {#2}, {#3}, #4}

\newcommand{\ybook}[3]{ #1, {#2} (#3)}

\newcommand{\tapj}[1]{ #1, {ApJ}, to be submitted}
\newcommand{\sapj}[2]{ #1, {ApJ}, submitted, arXiv:#2}

\usepackage{hyperref}
\usepackage[usenames,dvipsnames]{xcolor}

\colorlet{mylinkcolor}{violet}
\colorlet{mycitecolor}{Aquamarine}
\colorlet{myurlcolor}{YellowOrange}

\begin{document}

\title{The turbulent chiral magnetic cascade in the early universe}

\author{Axel~Brandenburg$^{1,2,3,4}$}
%\email{brandenb@nordita.org}

\author{Jennifer~Schober$^{3}$}
%\email{jschober@nordita.org}

\author{Igor~Rogachevskii$^{5,1,3}$}
%\email{gary@bgu.ac.il}

\author{Tina Kahniashvili$^{6,7}$}
%\email{tinatin@andrew.cmu.edu}

\author{Alexey~Boyarsky$^{8}$}
%\email{boyarsky@lorentz.leidenuniv.nl}

\author{J\"{u}rg~Fr\"{o}hlich$^{9}$}
%\email{juerg@phys.ethz.ch}

\author{Oleg~Ruchayskiy$^{10}$}
%\email{oleg.ruchayskiy@epfl.ch}

\author{Nathan~Kleeorin$^{5,3}$}
%\email{nat@bgu.ac.il}

\affiliation{
$^1$Laboratory for Atmospheric and Space Physics,
   University of Colorado, 3665 Discovery Drive, Boulder, CO 80303, USA \\
$^2$JILA and Department of Astrophysical and Planetary Sciences,
    Box 440, University of Colorado, Boulder, CO 80303, USA\\
$^3$Nordita, KTH Royal Institute of Technology
 and Stockholm University, Roslagstullsbacken 23,
 10691 Stockholm, Sweden \\
$^4$Department of Astronomy, AlbaNova University Center,
    Stockholm University, SE-10691 Stockholm, Sweden\\
$^5$Department of Mechanical Engineering,
 Ben-Gurion University of the Negev, P.O. Box 653, Beer-Sheva
 84105, Israel \\
$^6$McWilliams Center for Cosmology and Department of Physics,
  Carnegie Mellon University, 5000 Forbes Ave, Pittsburgh, PA 15213, USA\\
$^7$Abastumani Astrophysical Observatory, Ilia State University,
  3-5 Cholokashvili St., 0194 Tbilisi, Georgia\\
$^8$Instituut-Lorentz for Theoretical Physics, Universiteit Leiden,
 Niels Bohrweg 2, 2333 CA Leiden, The Netherlands\\
$^{9}$Institute of Theoretical Physics, ETH H\"{o}nggerberg,
 CH-8093 Zurich, Switzerland \\
$^{10}$Discovery Center, Niels Bohr Institute, Blegdamsvej 17,
 DK-2100 Copenhagen, Denmark
}

%\submitted{\today,~ $ $Revision: 1.223 $ $}
\submitted{Astrophys. J. 845, L21 (2017)}
\date{Received 2017 July 11; revised 2017 August 3; accepted 2017 August 8; published 2017 August 22}

\begin{abstract}
  The presence of asymmetry between fermions of opposite handedness in
  plasmas of relativistic particles can lead to exponential growth of a
  helical magnetic field via a small-scale chiral dynamo instability
  known as the chiral magnetic effect.  Here, we
  show, using dimensional arguments and numerical simulations, that this
  process produces through the Lorentz force chiral magnetically driven
  turbulence.  A $k^{-2}$ magnetic energy spectrum emerges via inverse
  transfer over a certain range of wavenumbers $k$.  The total chirality
  (magnetic helicity plus normalized chiral chemical potential) is conserved
  in this system.  Therefore, as the helical magnetic field grows, most of the
  total chirality gets transferred into magnetic helicity until the chiral
  magnetic effect terminates.
Quantitative results for height, slope, and extent of the spectrum are obtained.
Consequences of this effect for cosmic magnetic fields are discussed.
\end{abstract}


\section{Introduction}
\label{Introduction}

Asymmetry between the
number densities of left- and right-handed fermions gives rise to
what is known as the chiral magnetic effect (CME)
-- an electric current flowing along the magnetic field.
This quantum effect was first found by \citet{Vilenkin:80a}
and then rederived using different arguments
\citep{RW85,Alekseev:98a,Frohlich:2000en, Fukushima:08}.  \cite{JS97} and
\cite{Frohlich:2000en} showed that this phenomenon destabilizes a weak
magnetic field and leads to its exponential growth.
The CME has applications in many fields of physics ranging from the early
universe to neutron stars and condensed matter systems
\citep[for reviews, see, e.g.,][]{Kha14,2015PhR...576....1M}.

The total chirality in the system, i.e., the sum of magnetic helicity and
fermion chiral asymmetry,
is conserved. As the field becomes fully helical, the chiral asymmetry
will eventually disappear, so the total growth of magnetic fields
is limited \citep{JS97,Frohlich:2000en,BFR15,G13,TVV12,HKY15,PLS17}.

There is now significant interest in the possibility of generating a
turbulent inverse cascade by the CME
\citep{BFR12,BFR15,HKY15,DS17,PLS17}.
Meanwhile, there has been considerable progress in our understanding of
magnetically dominated helical turbulence \citep{BM99,KTBN13}.
In particular, the magnetic field $\BB$ decays with time $t$ such that
$\bra{\BB^2}\propto t^{-2/3}$ while
the correlation length grows like $\xiM\propto t^{2/3}$,
with $\bra{\BB^2}\,\xiM=\const$.
In this Letter we assess the importance of the CME in establishing initial
conditions for the turbulent decay.
This provides a critical starting point because we
predict the value of $\bra{\BB^2}\,\xiM$
based on the initial asymmetry.

It is worth noting that observational limits on the product $\bra{\BB^2}\,\xiM$
have been derived from the
non-observations of GeV-energy halos around TeV blazars \citep{Aharonian}.
This has been interpreted in terms of magnetic fields permeating the
intergalactic medium over large scales \citep[for a review, see][]{DN13}.
Simultaneous GeV--TeV observations of blazars put lower limits on such
fields between $10^{-15}\G$ \citep{TVN11} and
$10^{-18}\G$ \citep{Dermer} at $\xiM\sim 1\Mpc$.
Systematic parity-odd correlations between the directions of secondary photons
and their energies from the surroundings of blazars have been interpreted in
terms of helical magnetic fields of the order of $10^{-14}\G$
\citep{TCFV14,TV15}.
If this can be independently confirmed, it would be a real detection.

The present-day value of $\bra{\BB^2}\,\xiM$ and, more generally, the
modulus of the magnetic helicity can be constrained
on dimensional grounds under the assumption that it is determined
only by the present-day temperature $T_0$ plus fundamental
constants: the Boltzmann constant $\kB$,
the reduced Planck constant $\hbar$, and the speed of light $c$.
As the dimension of $\bra{\BB^2}$ is
$\erg\cm^{-3}=\G^2/4\pi$, we find
\EQ
\bra{\BB^2}\,\xiM=\epsilon(\kB T_0)^3(\hbar c)^{-2},
\label{argument}
\EN
where $\epsilon$ is a dimensionless number (we determine a more precise value
in Section~\ref{ApplicationEarlyUniverse}).  Assuming for now $\epsilon=1$ and
using $T_0=2.75\K$, \Eq{argument} yields the numerical value
$\sim5\times10^{-14}\erg\cm^{-2} =2\times10^{-37}\G^2\Mpc,$ corresponding to
$0.5\times10^{-18}\G$ at $1\Mpc$, and thus to the estimate of \cite{Dermer}.
Conversely, if this argument turned out to fail, it might suggest that other
dimensionful quantities such as Newton's constant might enter.
This argument is very
rough, and so quantitative models are needed to determine $\epsilon$.

The purpose of this Letter is to examine the onset of turbulence
by the chiral small-scale dynamo to compute the spectrum,
scale separation, and the saturation level of the resulting magnetic field.
We use three-dimensional simulations to verify scaling relations.
We conclude with a discussion of the parameters relevant to the
early universe and, in particular, the strength of the observable
magnetic field.

\section{Growth and saturation}
\label{GrowthSaturation}

The CME can lead to exponential growth owing
to a term in the induction equation that is formally similar to
the $\alpha$ effect in mean-field dynamo theory \citep{Mof78,KR80}.
The important difference is of course that in chiral magnetohydrodynamics we
are concerned with the
actual magnetic field rather than its macroscopic average.
However, much of the intuition from mean-field electrodynamics carries
over to chiral magnetohydrodynamics.
The mathematical formalism and the underlying fully nonlinear
spatially dependent evolution equations were derived by \cite{BFR15}
and analyzed by \cite{Roga17}.

Since $\BB$ is solenoidal, we express it as $\BB=\nab\times\AAA$.
We define $\mu=24\,\alphaem\,(n_{\rm L}-n_{\rm R})\,(\hbar c/\kB T)^2$
as the normalized chemical potential,
where $\alphaem\approx1/137$ is the fine structure constant,
and $n_{\rm L}$ and $n_{\rm R}$ are the number densities
of left- and right-handed fermions, respectively.
The governing equations for $\AAA$ and $\mu$ are
\EQ
{\partial\AAA\over\partial t}=\eta\left(\mu\BB-\nab\times\BB\right)+\UU\times\BB,
\label{dAdt}
\EN
\EQ
{\DD\mu\over\DD t}=-\lambda\,\eta\left(\mu\BB-\nab\times\BB\right)\cdot\BB
+D\nabla^2\mu-\Gamma_{\rm\!f}\mu,
\label{dmudt}
\EN
where $\DD/\DD t\equiv\partial/\partial t+\UU\cdot\nab$ is the
advective derivative, $\eta$ is the magnetic diffusivity
(not to be confused with the conformal time\footnote{Our evolution
equations are also valid in an expanding universe when interpreting $t$
as conformal time and using comoving quantities; see \cite{BEO96}, whose
equations contain extra terms and $4/3$ factors that affect $\rho$ and
$\uu$ only slightly and do not affect our results.}),
$\lambda\equiv{}3\hbar c\,(8\alphaem/\kB T)^2$
characterizes the feedback of the electromagnetic field
on the evolution of $\mu$, $D$ is a chiral diffusion coefficient,
$\Gamma_{\rm\!f}$ is the flipping rate,
and $\UU$ is the plasma velocity, which obeys the usual momentum equation
and the continuity equation for the density $\rho$,
\EQ
\rho{\DD\UU\over\DD t}=(\nab\times\BB)\times\BB-\nab p
+\nab\cdot(2\rho\nu\SSSS),
\label{dUdt}
\EN
\EQ
{\DD\rho\over\DD t}=-\rho\nab\cdot\UU,
\label{drhodt}
\EN
where ${\sf S}_{ij}=\half(U_{i,j}+U_{j,i})-\onethird\delta_{ij}\nab\cdot\UU$
is the rate-of-strain tensor, $\nu$ is the viscosity, and $p$ is the
pressure, which is assumed to be proportional to the density, i.e.,
$p=\rho\cs^2$, with $\cs$ being the speed of sound;
for a gas of ultra-relativistic particles, $\cs^2=c^2\!/3$.

The magnetic field is normalized such that the magnetic energy
density is $\BB^2/2$ without the $4\pi$ factor.
The usual magnetic field in Gauss is therefore $\sqrt{4\pi}\,\BB$.
Furthermore, if $\mu=\lambda=0$, we recover the usual hydromagnetic
equations.
The physical values of $\lambda$ and $\eta$, as well as the initial values
of $\mu$, will be discussed in \Sec{ApplicationEarlyUniverse} in the
context of the early universe.
In the following, however, we consider a broad parameter space and
are particularly interested in the case where the magnetic field can
grow out of a weak seed due to the CME.

An important consequence of \Eqs{dAdt}{dmudt} is the conservation
of (volume-averaged) total chirality:
\EQ
\half\lambda \, \bra{\AAA\cdot\BB}+\bra{\mu}=\const\equiv\mu_0
\quad\mbox{(for $\Gamma_{\rm f}\ll\eta\mu_0^2$)},
\label{totchirality}
\EN
where $\mu_0$ denotes the initial total chirality and the
brackets denote averaging over a closed or periodic volume.%
\footnote{At high temperatures we neglect the influence of the fermions' mass
that would gradually destroy the total chirality~(\ref{totchirality}).}
\EEq{totchirality} imposes an important constraint on the coevolution of $\mu$
and $\BB$, and implies
\EQ
\bra{\BB^2}\,\xiM\la\mu_0/\lambda,
\label{B2sat}
\EN
where $\xiM$ is the correlation length
discussed in \Sec{Introduction}.

The initial growth can be described by the linearized equations.
Assuming $\BB(t,\xx)\propto\exp(\gamma t+\ii\kk\cdot\xx)$, where $\gamma$ is
the dynamo growth rate
and $\kk$ is the wavevector, the dispersion relation becomes
$\gamma(k)=\eta k\,(\mu-k)$, where $k=|\kk|$ is the wavenumber.
This dispersion relation predicts magnetic field growth for $k<\mu$.
The linear approximation is applicable as long as the field is weak
and nonlinear effects are small.
In a domain of size $L$, which could represent the horizon scale
in the early universe or the length of a periodic computational domain,
there will be a minimum wavenumber $\kone =2\pi/L$.
A weak seed magnetic field is only unstable if $\kone <\mu$.
In addition, there is a wavenumber $k_{\max}$ at which the growth rate
is maximum, i.e., $\dd\gamma/\dd k=0$.
It is given by $k_{\max}=\mu/2$.
Eventually, following the early exponential growth, the magnetic field
reaches a critical value and a more complicated nonlinear stage commences.

The CME introduces two new quantities into the
system: $\lambda$ and $\mu$.
Different evolutionary scenarios can be envisaged depending on their values.
For the purpose of this discussion, we ignore the fact that $\mu$ is
changing as $\BB$ evolves; all normalizations below are actually based
on $\mu_0$.
Using the fact that
$\lambda^{-1}$ has the dimension of energy per unit length
and $\mu$ has the dimension of inverse length, we can identify
two characteristic velocities:
\EQ
\vsat =\mu/(\meanrho\lambda)^{1/2},\quad\quad
v_\mu=\mu\eta,
\EN
where $\meanrho$ is the mean density of the plasma.
Assuming that both velocities are well below the speed of sound, $\cs$,
we can identify two regimes of interest:
\EQ
\cs > \vsat  > v_\mu > \eta \kone  \quad\mbox{(regime I)},
\label{regI}
\EN
\EQ
\cs > v_\mu > \vsat  > \eta \kone  \quad\mbox{(regime II)}.
\label{regII}
\EN
In regime~I, the ratio $\vsat/v_\mu=[\eta\,(\meanrho\lambda)^{1/2}]^{-1}$
is large, so the $\lambda$ term is unimportant and $\mu$ will only change slowly
as the magnetic field grows.
Once the magnetic field exceeds a critical value of around
$\sqrt{\meanrho}\mu\eta$, i.e., when
the magnetic Reynolds number exceeds a certain value,
the magnetic field becomes turbulent.
However, the field continues to grow until
the Alfv\'{e}n speed $\vA=B/\sqrt{\meanrho}$ approaches $\vsat$.
At that point, $\mu$ begins to be depleted, as is clear from
\Eq{totchirality}.
This quenches further growth, and magnetic energy can then only decay.
In regime~II, the ratio $\vsat/v_\mu$ is small, so the $\lambda$ term
is important and $\mu$ will be depleted before a turbulent cascade develops.
As in regime I, the magnetic energy must eventually decay.
In both cases, however, since the magnetic field is
maximally helical, the decay will be slower than for a nonhelical
field and will be accompanied by strong, self-similar inverse transfer
\citep{CHB01,BK17}.

\begin{figure}[t!]\begin{center}
\includegraphics[width=\columnwidth]{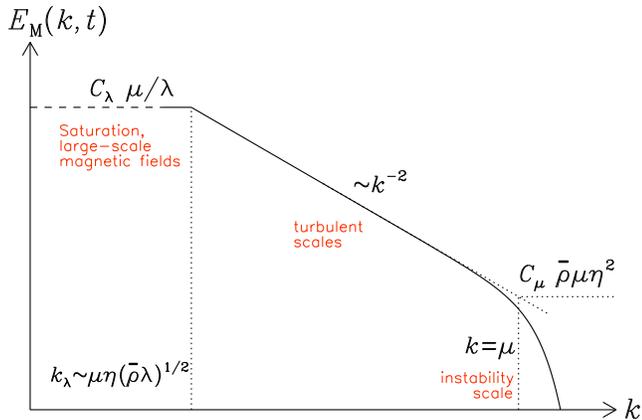}
\end{center}\caption[]{Sketch of the magnetic energy spectrum of
  chiral magnetically driven turbulence.
}\label{psketch}
\end{figure}

In the following, we will see that a more accurate distinction between
regimes~I and II occurs when $\vsat/v_\mu\approx8$ instead of unity.
For this purpose, let us discuss the resulting
magnetic spectrum in more detail.
The spectrum is defined such that
$\int E_{\rm M}(k,t)\,\dd k =\bra{\BB^2}/2$.
Thus, $E_{\rm M}(k,t)$ has the dimension of $\meanrho\mu\eta^2$, so
it will be convenient to normalize $E_{\rm M}(k,t)$ correspondingly.
In magnetically dominated turbulence, where the velocity is just a
consequence of driving by the Lorentz force, we expect weak turbulence
scaling \citep{Gal00} with a spectrum proportional to $k^{-2}$.
This was originally thought to be applicable to the case with an
external magnetic field, but this spectrum was found also
for the isotropic case of magnetically dominated turbulence \citep{BKT15}.
On dimensional grounds, we can
expect a spectrum with an inertial range of the form
\EQ
E_{\rm M}(k,t)=C_\mu\,\meanrho\mu^3\eta^2k^{-2},
\label{Cmu}
\EN
where $C_\mu$ is a chiral magnetic Kolmogorov-type constant.
Note that $\lambda$ does not enter in \Eq{Cmu}.
This cannot be justified by dimensional arguments alone and requires
verification from simulations that will be presented below.
It should be noted, however, that $\lambda$ characterizes
the ultimate depletion
of $\mu$ after turbulence develops and the magnetic field saturates.

The depletion of $\mu$, and therefore of the CME, becomes
stronger with increasing values of $\lambda$, thus limiting the
magnetic field to progressively smaller values as $\lambda$ is increased.
Using \Eq{B2sat}, we find that
\EQ
E_{\rm M}(k,t)\leq C_\lambda\,\mu/\lambda,
\label{Clam}
\EN
where $C_\lambda$ is another Kolmogorov-type constant describing
the saturation caused by $\lambda$.
Note that this limit for $E_{\rm M}(k,t)$ is independent of $k$ and
applicable to all $k\leq k_\lambda$, where $k_\lambda$ is a
critical value that will be estimated below.

\EEq{Clam} suggests that in regime~I, we have an inertial range
for $E_{\rm M}(k,t)<C_\lambda\,\mu/\lambda$ with a $k^{-2}$
spectrum down to the smallest excited scale
at wavenumber $k=\mu$.
Using Equations \eqs{Cmu}{Clam}, we can determine a critical value
$k_\lambda$ as
\EQ
k_\lambda=\sqrt{\meanrho\lambda C_\mu/C_\lambda}\,\mu\eta.
\label{klambda}
\EN
The resulting $k^{-2}$ spectrum in $k_\lambda<k<\mu$ is sketched in
\Fig{psketch}.
It is now clear that the crossover between regimes~I and II occurs when
$k_\lambda\approx\mu/2$, i.e., when
$\vsat/v_\mu\approx2\sqrt{C_\mu/C_\lambda}$.
To determine the constants $C_\mu$ and $C_\lambda$, we need
numerical simulations, which will be presented in the next section.

\begin{figure*}[t!]\begin{center} %(1., new)
\includegraphics[width=.49\textwidth]{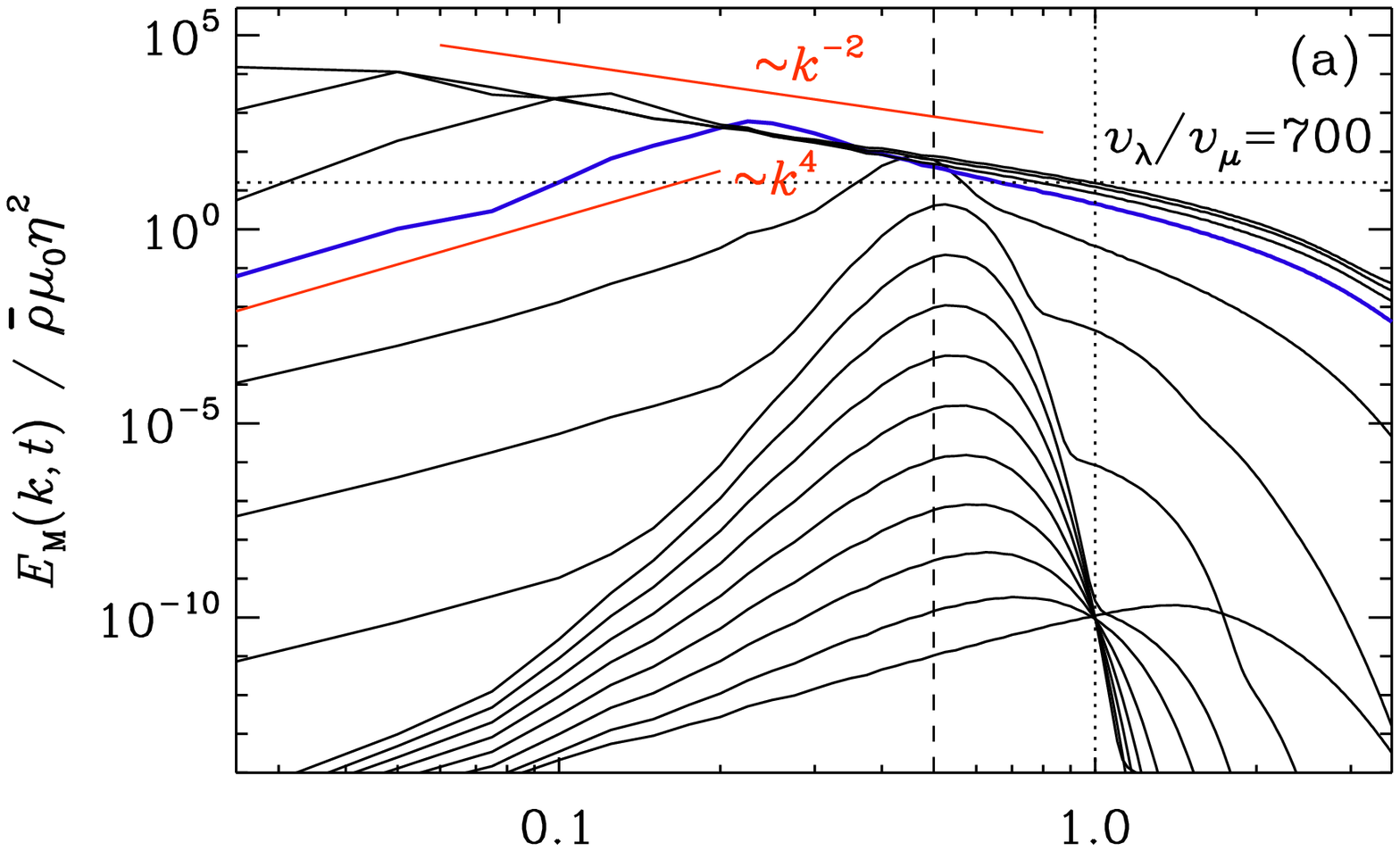}
\includegraphics[width=.49\textwidth]{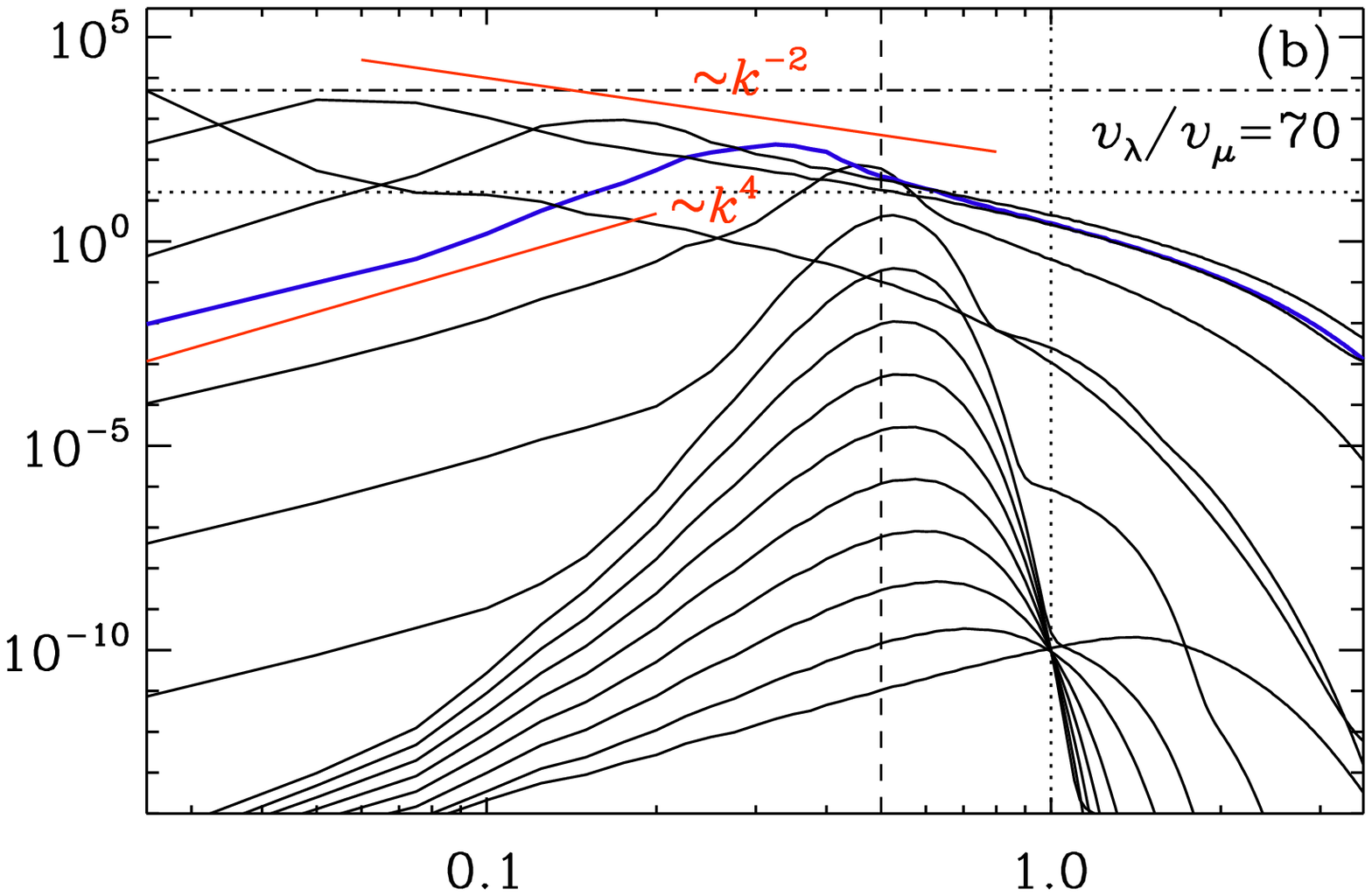}
\includegraphics[width=.49\textwidth]{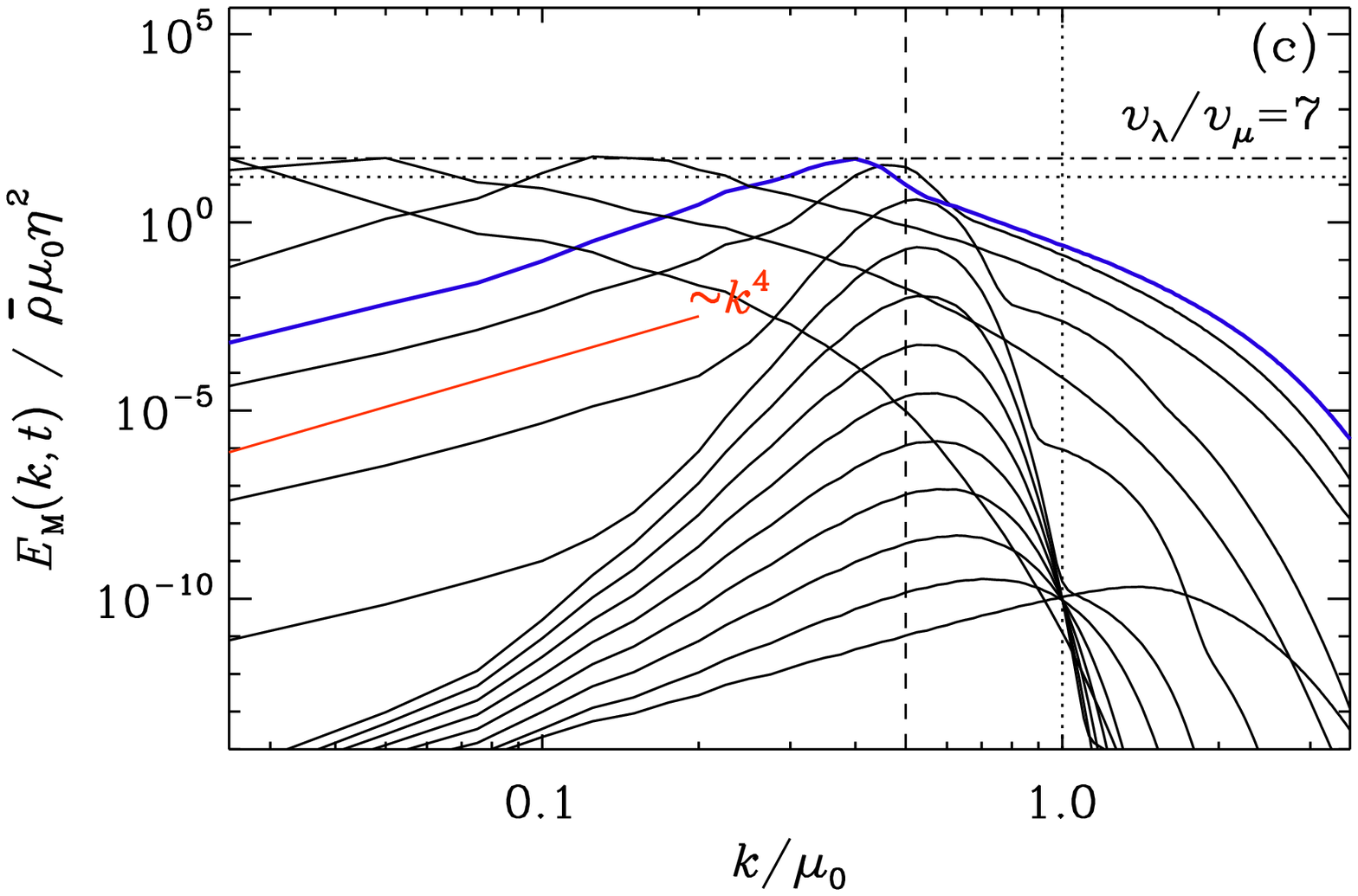}
\includegraphics[width=.49\textwidth]{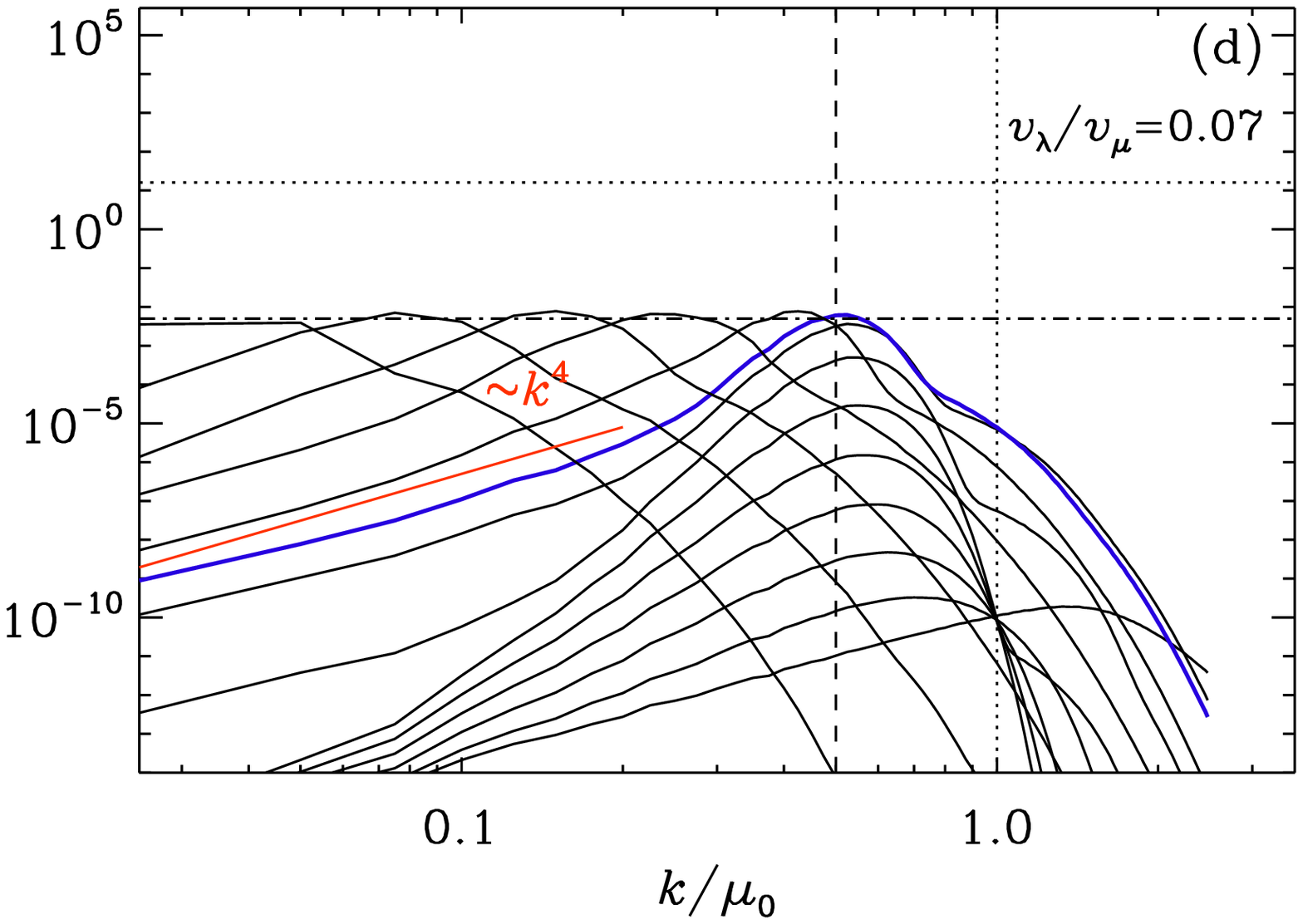}
\end{center}\caption[]{
Evolution of magnetic energy spectra for Runs A--D shown in time intervals
$\Delta t$ with $\Delta t\,\eta\mu^2=6$ until $t\,\eta\mu^2=80$ (marked in
blue) followed by longer time intervals that increase by a factor of two.
The vertical dashed line marks the wavenumber $k=\mu_0/2$
where the growth rate is maximum.
The horizontal dotted line marks the level of $C_\mu\meanrho\mu_0\eta^2$,
and the horizontal dashed-dotted line marks the level of $C_\lambda\mu_0/\lambda$.
The red lines have slopes of $-2$ and $+4$.
Panel (d) corresponds to regime~II, while panels (a) and (b) correspond to regime~I.
}\label{pspec_select}\end{figure*}

\section{Simulations}
\label{sec:simulations}

Below, we present results of
three-dimensional simulations in a cubic periodic domain.
Detailed analysis of different simulations with
chiral magnetically driven turbulence and
externally forced turbulence are presented in a separate
study \citep{Schober17}.
We use the {\sc Pencil Code}\footnote{\url{https://github.com/pencil-code}}
with $288^3$ and $200^3$ mesh points.
We vary the values of $\mu$, $\eta$, and $\lambda$, covering both
regimes~I and II.
Numerical stability requires $\nu$ and $D$ to be larger than what
is physically realistic, so we take $\nu=D=\eta$ as a compromise.
As in earlier work, we assume $\Gamma_{\rm\!f}=0$.
The governing parameters are listed in \Tab{Tsum}.

In \Fig{pspec_select} we show magnetic energy spectra
for different values of $\vsat /v_\mu$.
As explained in \Sec{GrowthSaturation}, the depletion of $\mu$ is small
if $\lambda$ is small and thus $\vsat $ is large, so for
$\vsat /v_\mu\gg1$, a turbulent cascade with a power law as in
\Eq{Cmu} is possible.
This is shown in panel (a), where $\vsat /v_\mu=700$, and a $k^{-2}$
spectrum is seen for all $k<\mu$.
As $\vsat /v_\mu$ is decreased, depletion of $\mu$ is increased;
see \Figp{pspec_select}{b}.
The limiting line where $E_{\rm M}(k,t)=\mu/\lambda$, is shown as a
dashed-dotted line, and we see that a typical inverse transfer sets in
as found previously for decaying turbulence \citep{CHB01,BJ04,BK17}.

Next, we discuss the values of $C_\lambda$ and $C_\mu$.
\Figsp{pspec_select}{c}{d} have already demonstrated that
$C_\lambda\approx1$ is a good approximation.
The value of $C_\mu$ can be seen from the intercept of the $k^{-2}$
power law with the $k=\mu$ line in \Figsp{pspec_select}{a}{b}.
We see that the intercept lies at $C_\mu\approx16$.
This implies that the crossover between regimes~I and II is at
$\vsat/v_\mu\approx2\sqrt{C_\mu}\approx8$, which is compatible
with \Figp{pspec_select}{c}.

\begin{figure*}[t!]\begin{center} %(1., new)
\includegraphics[width=.24\textwidth]{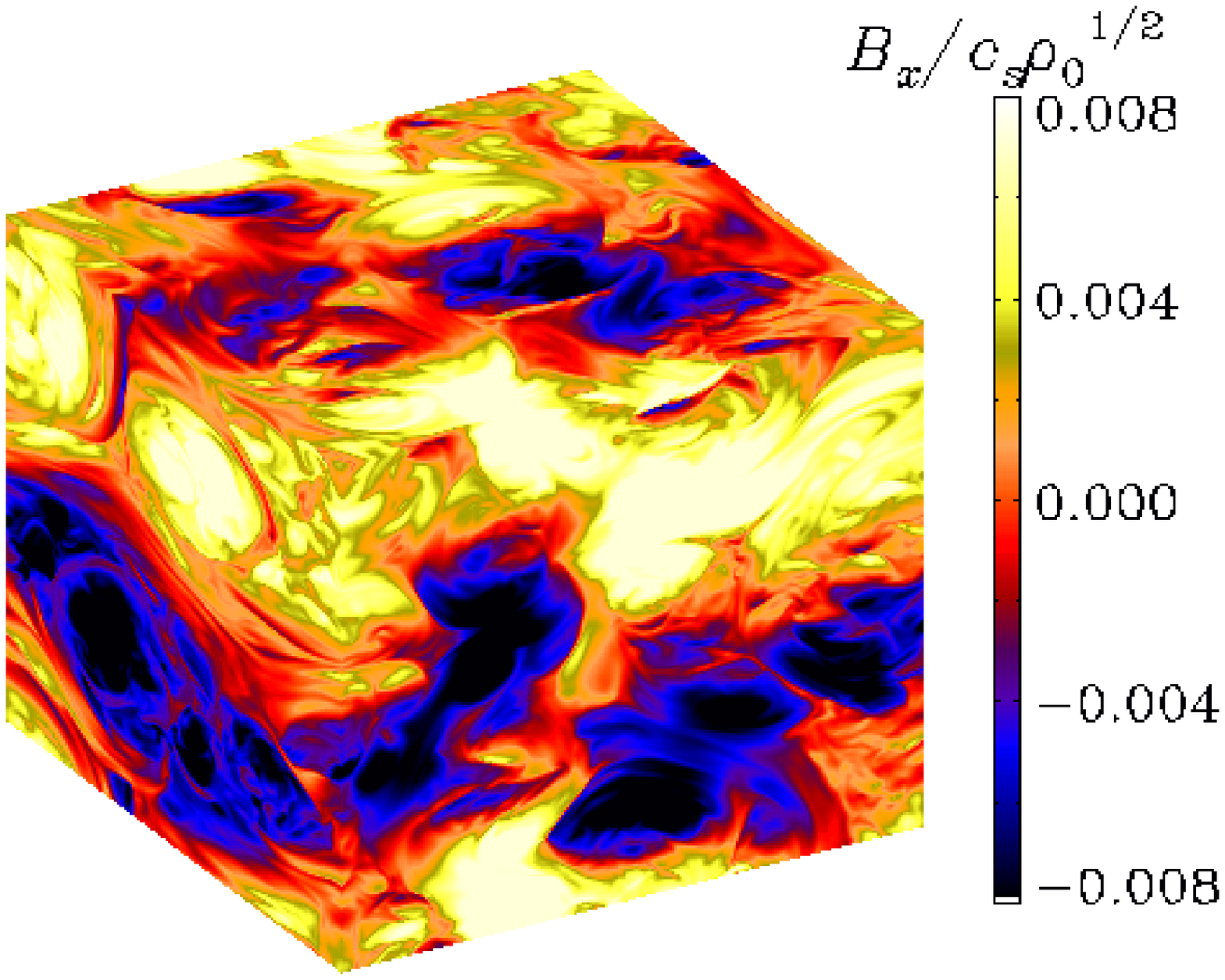}
\includegraphics[width=.24\textwidth]{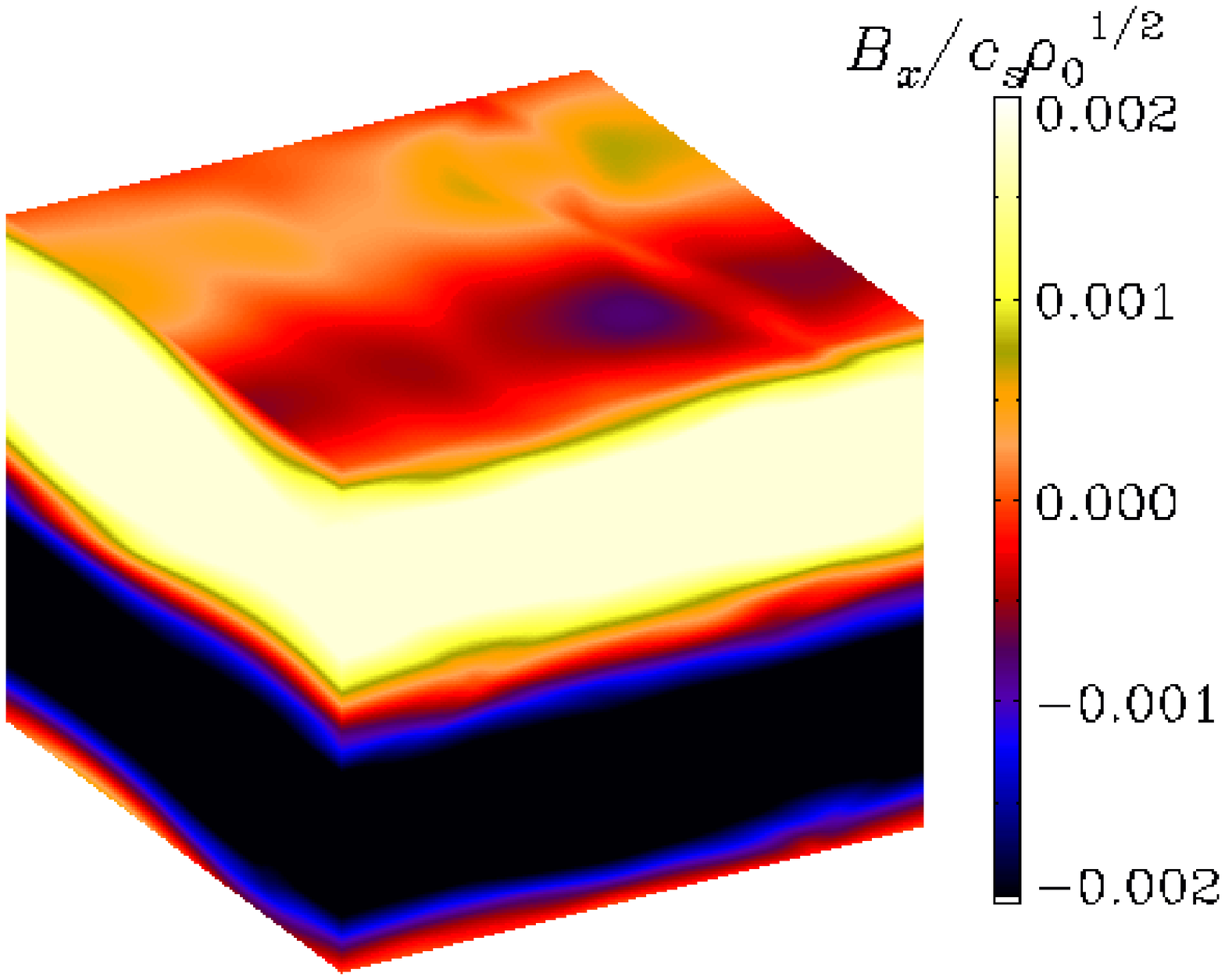}
\includegraphics[width=.24\textwidth]{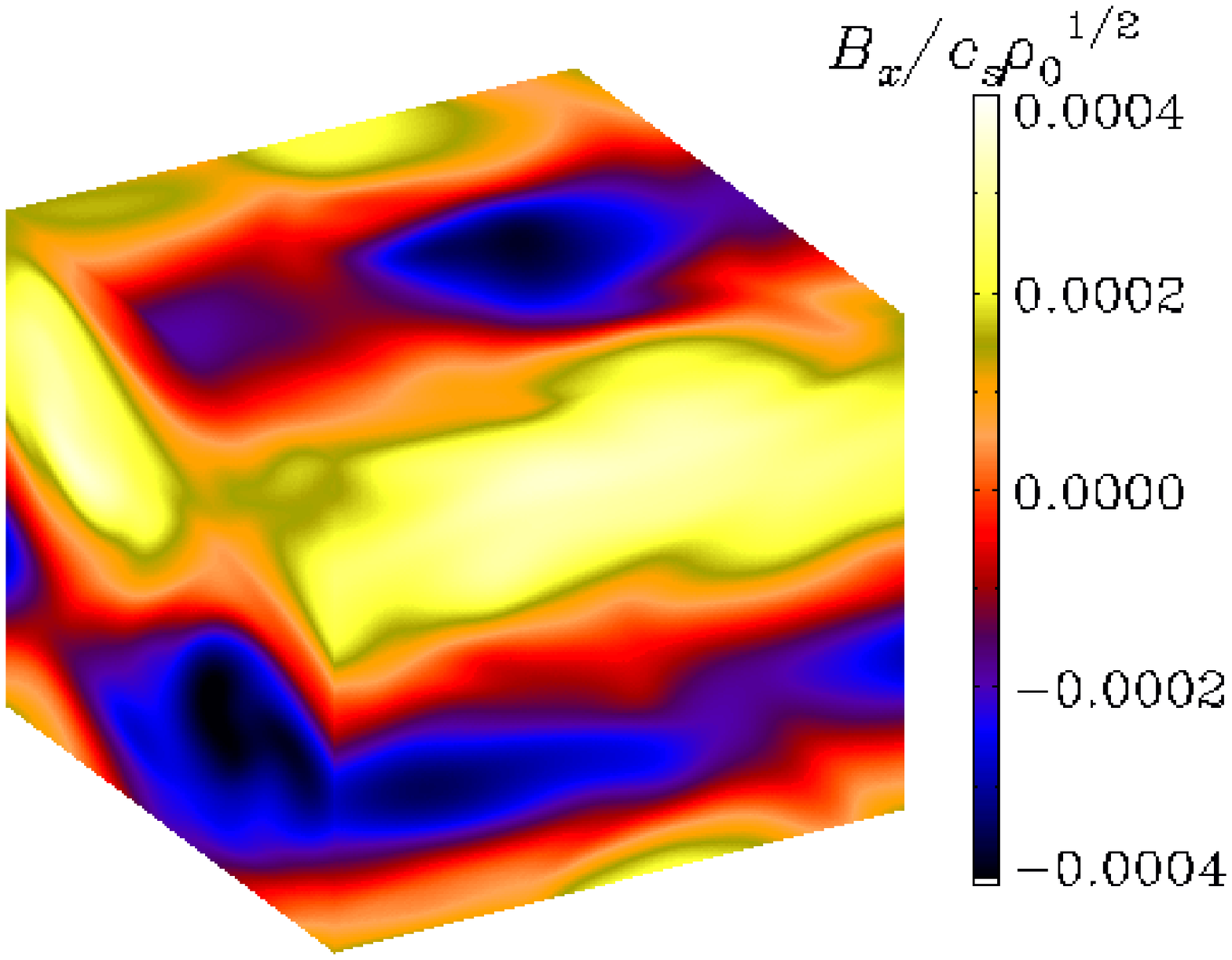}
\includegraphics[width=.24\textwidth]{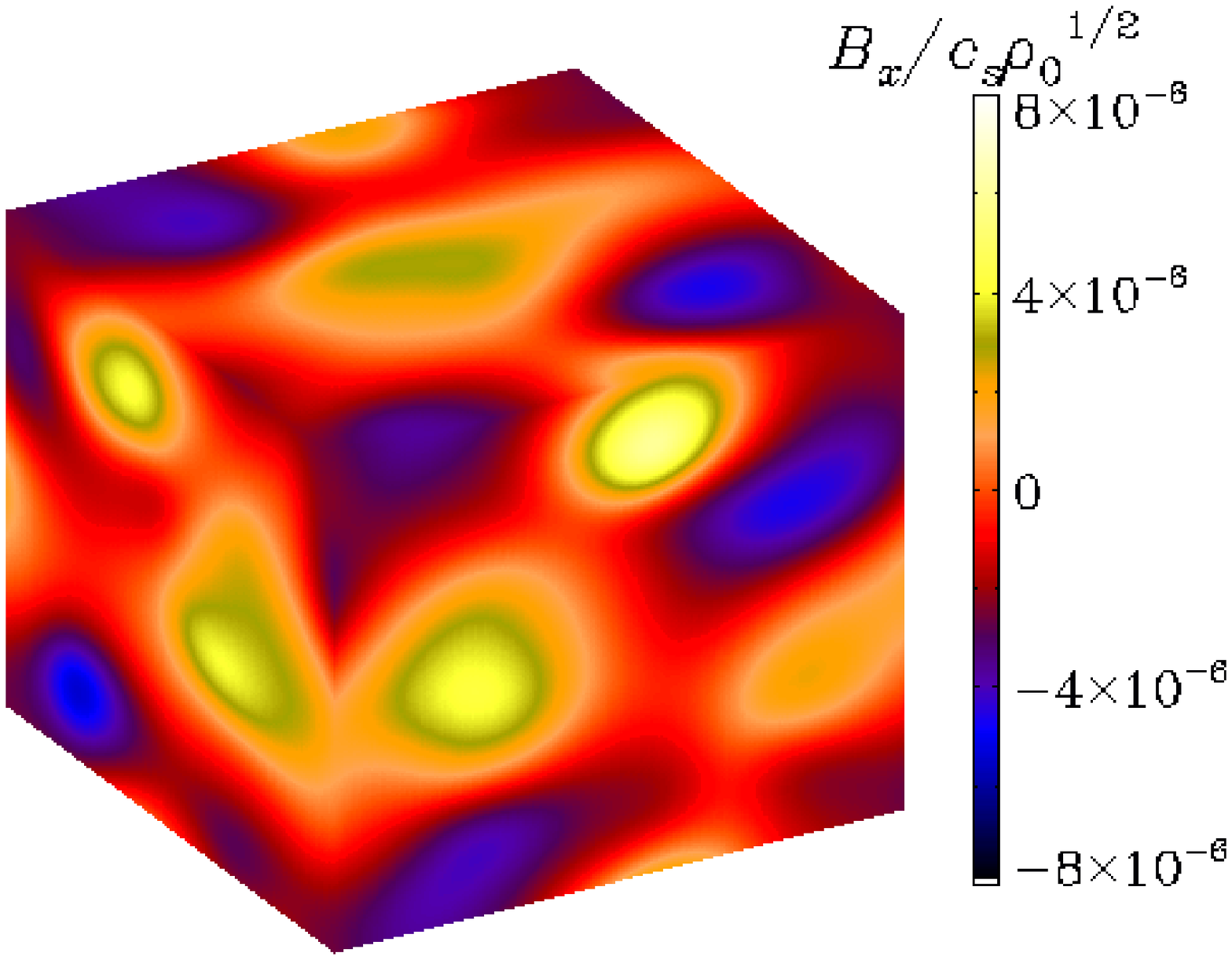}
\includegraphics[width=.24\textwidth]{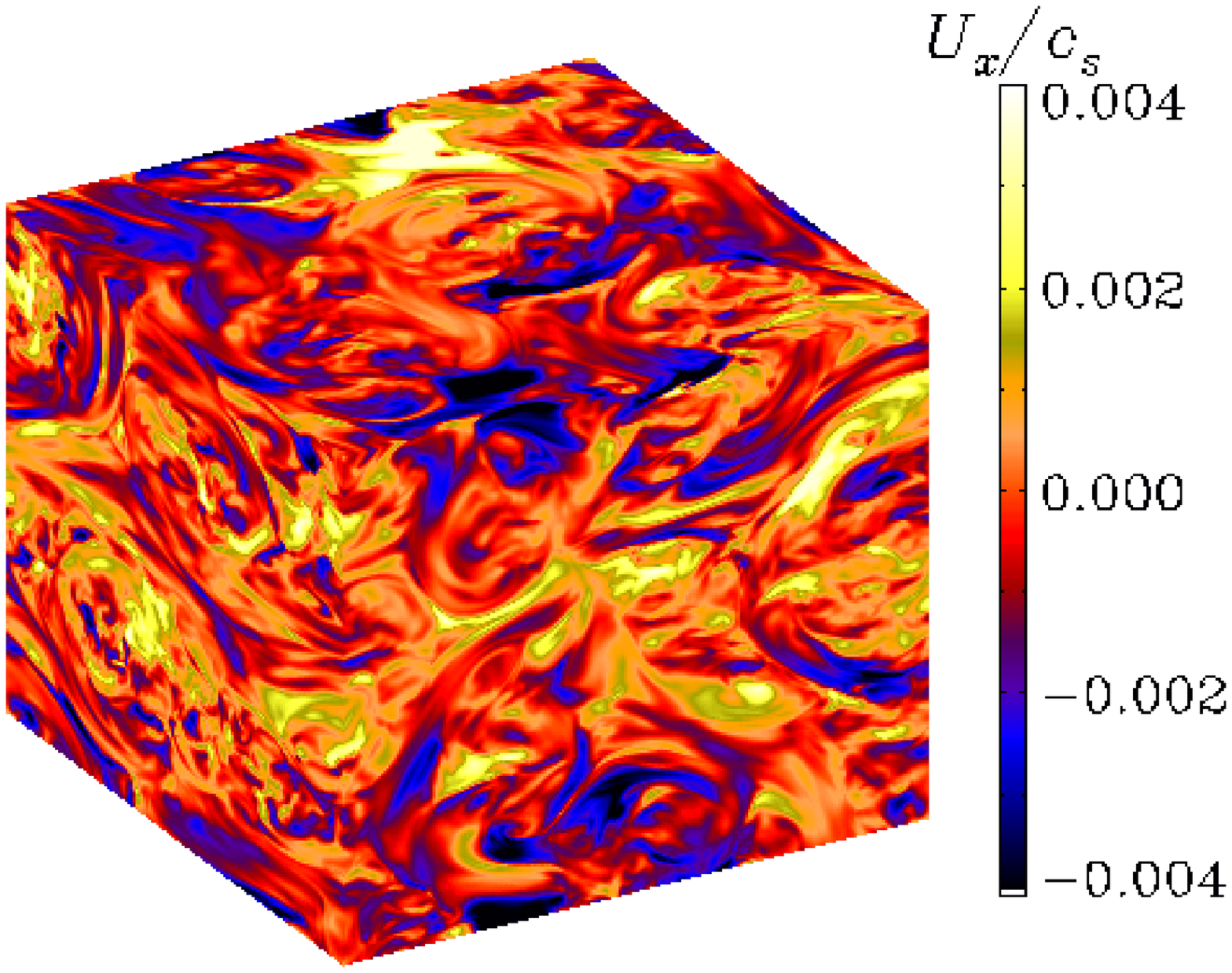}
\includegraphics[width=.24\textwidth]{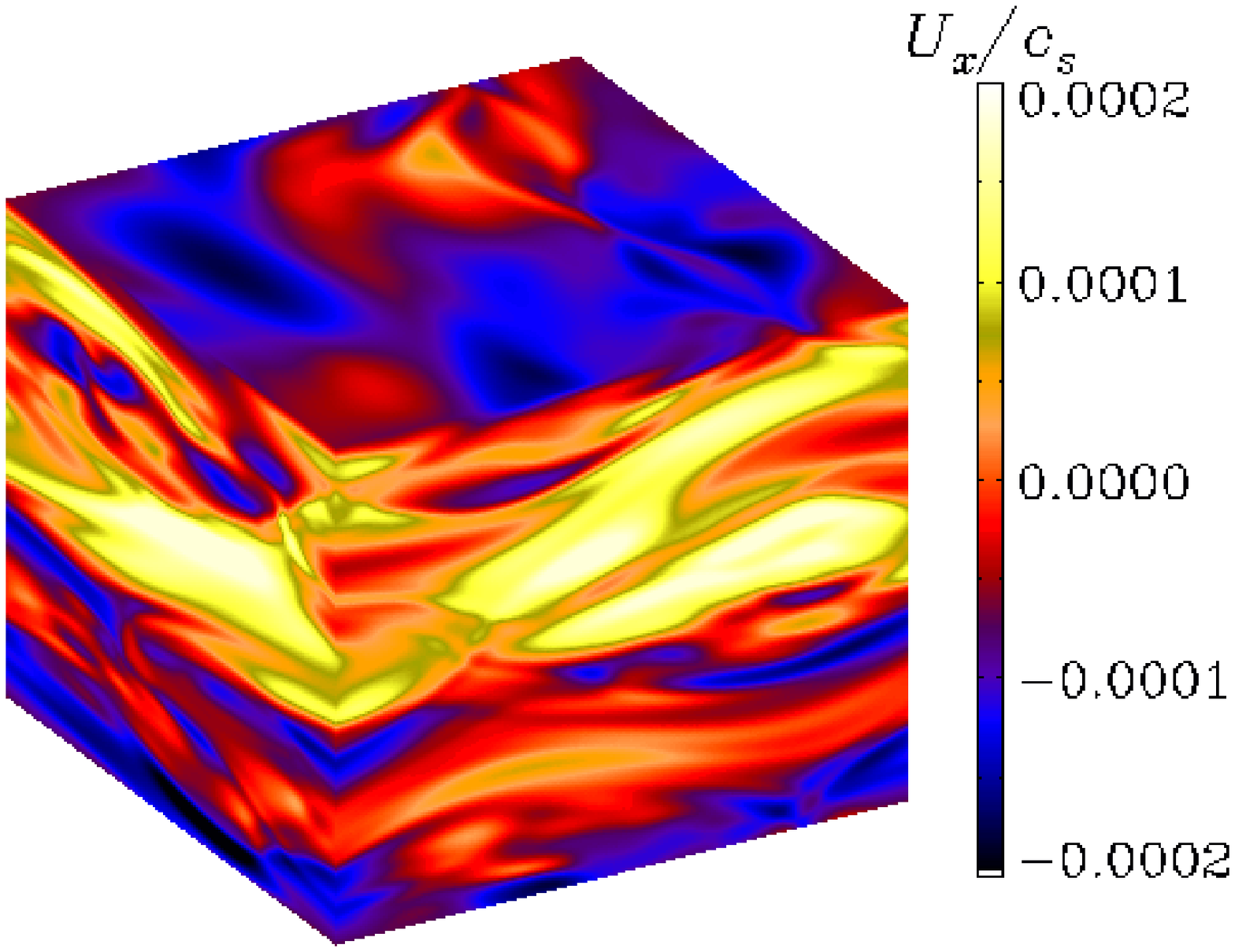}
\includegraphics[width=.24\textwidth]{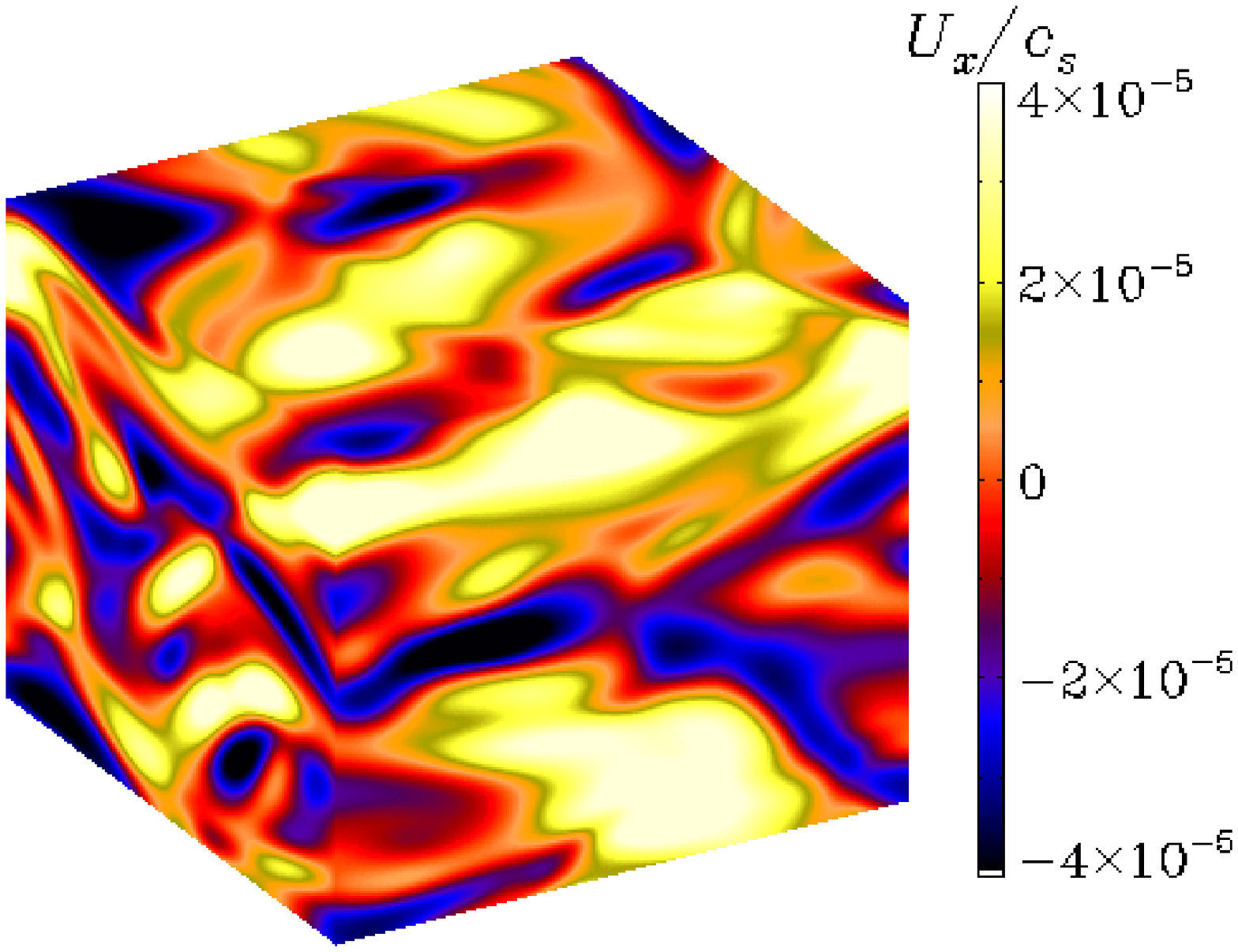}
\includegraphics[width=.24\textwidth]{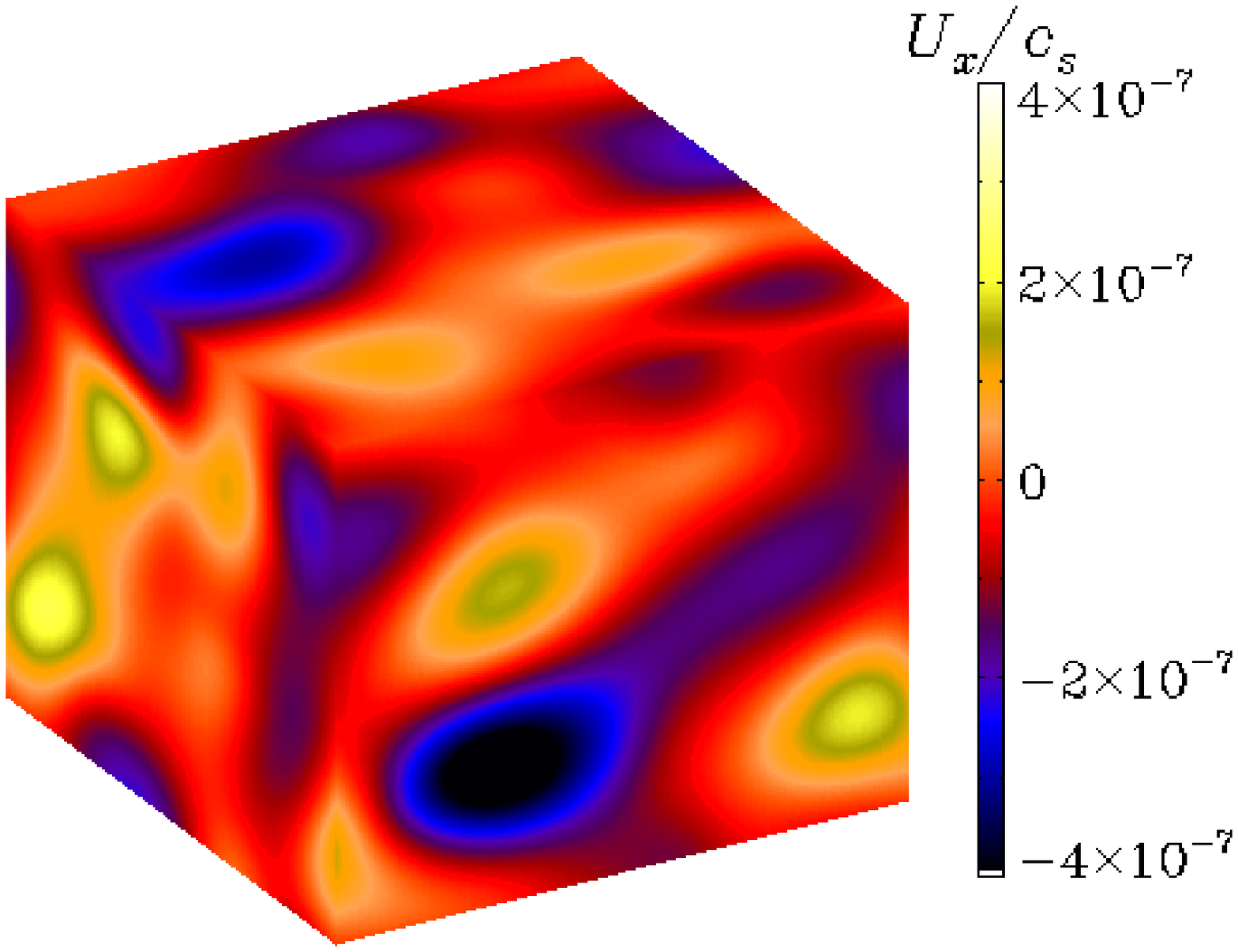}
\end{center}\caption[]{
$B_x$ and $U_x$ on the periphery of the computational domain for
(from left to right) $\vsat /v_\mu=700$, $70$, $7$, and $0.07$
at the last time.
\vspace{12mm}
}\label{psnap}\end{figure*}

For $\vsat /v_\mu<8$, the inverse transfer begins once the
approximately monochromatic exponential growth at $k=\mu/2$ saturates.
Both for large and small values of $\vsat /v_\mu$, the magnetic field
is turbulent, as shown in \Fig{psnap}, where we compare visualizations
of $B_x$ and $U_x$ on the periphery of the computational domain.
$\BB$ attains a large-scale component of Beltrami-type,
which is force-free and of the form $(\sin \kone z,\cos \kone z, 0)$
with positive helicity.
It is a matter of chance in which direction the field varies.
Examples of fields varying in any of the other two directions have
been found for helically forced turbulent dynamos \citep{Bra01}.

Initially and at late times, $\mu$ is nearly uniform.
At intermediate times, however, the ratio of its rms value to the
average can reach 25\% in the case with $\vsat /v_\mu=7$.
The typical kinetic energy can reach 12\% of the magnetic energy.

\begin{figure}[t!]\begin{center}
\includegraphics[width=\columnwidth]{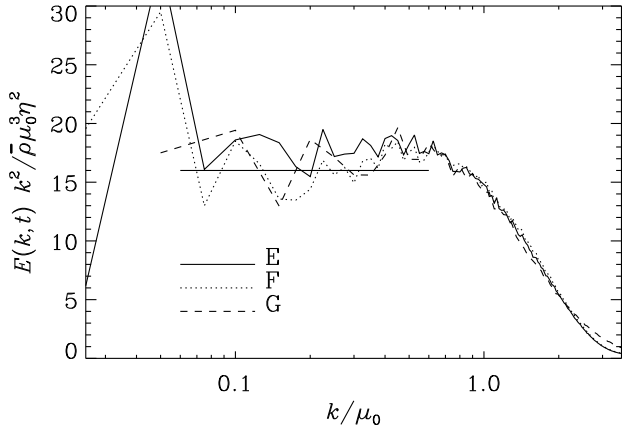}
\end{center}\caption[]{
Compensated spectra for Runs E--G.
}\label{ppspec_comp}\end{figure}

The value of $C_\mu$ can be determined more accurately by plotting energy
spectra compensated by $k^2/\mu^3\eta^2$; see \Fig{ppspec_comp}.
We see that $C_\mu\approx16$ is well obeyed for different values of
$v_\mu/\cs$ and $\mu_0/\kone $; see \Tab{Tsum}.
Here, we also give the Lundquist number $\Lu=\vA/\eta \kone $
and the Reynolds number $\Rey=\urms/\nu \kone $ at the end of the run.
By integrating \Eq{Cmu} over $k$, we obtain the estimate
\EQ
\Lu={\mu\over\kone}\left({2\,C_\mu\,\mu\over k_\lambda}\right)^{1/2}
=(C_\mu C_\lambda)^{1/4} {\mu\over\kone}
\left({2\,v_\lambda\over v_\mu}\right)^{1/2}.
\EN
This is compatible with the values in \Tab{Tsum}, especially
for Run~B, while for other runs they are only lower limits.

\begin{table}[b!]\caption{
Summary of the parameters for the runs discussed.
}\vspace{12pt}\centerline{\begin{tabular}{cccccccc}
$\!\!$Run$\!\!$& $\mu_0$ & $\eta\kone/\cs$ & $\meanrho\lambda\cs^2/\kone ^2$ &$v_\mu/\cs$&$\vsat /\cs$&$\Lu$&$\Rey$ \\
\hline
A& 40    &$5\times10^{-5}$ &$8\times10^2$ &$0.002$&    1.4    &$\!\!2000^*\!\!\!$&$\!\!450^*\!\!\!$\\ %288_3D_kf60_mu040_lambda1e3d_double
B& 40    &$5\times10^{-5}$ &$8\times10^4$ &$0.002$&   0.14    &$  830$&$ 250$\\ %288_3D_kf60_mu040_lambda1e5d_double2
C& 40    &$5\times10^{-5}$ &$8\times10^6$ &$0.002$&  0.014    &$  170$&$  47$\\ %288_3D_kf60_mu040_lambda1e7d_double
D& 40    &$5\times10^{-5}$&$8\times10^{10}\!\!$&$0.002$&$\!\!1.4\times10^{-4}\!\!$&$1.5$&$0.05$\\ %288_3D_kf60_mu040_lambda1e11d_double
E& 40    &$5\times10^{-5}$&$       10^2  $&$0.002$&     4.0   &$2000^*\!\!\!$&$ 450$\\ %288_3D_kf60_mu040_lambda1e2d
F& 40    &$5\times10^{-6}$&$       10^2  $&$0.0002$&    4.0   &$2500^*\!\!\!$&$ 500^*\!\!\!$\\ %288_3D_kf60_mu040_lambda1e2c
G& 20    &$5\times10^{-6}$&$       10^2  $&$0.0001$&    2.0   &$ 800^*\!\!\!$&$ 200$\\ %288_3D_kf60_mu020_lambda1e2c
\label{Tsum}
\end{tabular}}
\tablenotemark{$\mu_0$ is in units of $\kone$. Asterisks denote lower limits.}
\end{table}

The early exponential growth of the magnetic field is superficially
reminiscent of a turbulent dynamo.
A major difference is, however, the absence of an initial transient,
which is usually found in dynamo simulations \citep{Bra10}.
In all cases, we started with a random magnetic field with a Batchelor
$k^4$ spectrum \citep{DC03}.
Looking again at \Figssp{pspec_select}{a}{d}, we see that at
later times the field depends only on the initial value of $E_{\rm M}(\mu/2,0)$
and is independent of the spectral energy at any other $k$,
so we have
\EQ
E_{\rm M}(\mu/2,t)=E_{\rm M}(\mu/2,0)\,\exp(\gamma_{\max} t),
\EN
where $\gamma_{\max}=\eta\mu^2/4$ is the value of $\gamma(k)$
at $k=\mu/2$.
At $k=\mu_0$, the spectral energy is strictly constant
in the kinematic phase, so all lines cross at that one point,
where the spectral magnetic energy density is equal to the initial value.
The magnetically driven chiral turbulence and resulting large-scale
magnetic fields are similar to cosmic ray-driven
turbulence through the Bell instability \citep{RKBE12}.

\section{Application to the early universe}
\label{ApplicationEarlyUniverse}

Our work has shown that the CME acts as an intermediary
to the previously studied cases of decaying hydromagnetic turbulence.
It yields the initial conditions:
the length scale, $\xiM=k_\lambda^{-1}$, and the field strength,
$|\BB|\approx(\mu k_\lambda/\lambda)^{1/2}$; see \Eq{B2sat}.

We now apply our findings to the case of the cosmological magnetic field.
The equations of \Sec{GrowthSaturation} remain unchanged if one uses
\emph{comoving} variables \citep[see][]{Roga17}.

The initial value of chiral asymmetry depends on microscopic physics.
To establish a model-independent upper limit,
it is instructive to write the conservation of total chirality
of \Eq{totchirality}, using the definitions of $\mu$ and $\lambda$
of \Sec{GrowthSaturation}, in the form
\begin{equation}
  \label{eq:1}
  (n_{\rm L} - n_{\rm R}) + \frac{4\alphaem}{\hbar c}
  \langle \AAA\cdot \BB\rangle = \const.
\end{equation}
Assuming that fermions are in equilibrium with photons,
the maximal value of chiral asymmetry in one fermion species is attained when
$n_{\rm L} \approx n_\gamma$---the number of photons---and
$n_{\rm R} \approx 0$.
In this case, the maximal value of magnetic helicity today would be
\begin{equation}
  \label{eq:3}
  \bra{\BB^2}\xiM
    =\frac{\hbar c}{4\alphaem} \frac{g_{0}}{g_\ast} n_{\gamma0} N_{\rm f}
    = 5\times 10^{-38} \frac{N_{\rm f}}{10}g_{100}^{-1} \G^2\Mpc.
\end{equation}
Here, $g_0=3.36$ and
$n_{\gamma0} = 2\zeta(3)/\pi^2 (k_B T_0/\hbar c)^3 = 411\cm^{-3}$ is the
number of photons today, the factor $N_{\rm f}$ takes into account that many
relativistic fermions with asymmetric populations are present in the plasma,
and therefore the total fermion chirality exceeds $n_\gamma$,
$g_{100}=g_*/100$ is the
effective number of degrees of freedom at temperatures where magnetic fields
are generated (in the Standard Model
$g_*=106.75$ at $T \sim 100\GeV$), and $\zeta(3)\approx1.202$.
If all fermions have comparable asymmetries at high temperature, the
estimate~(\ref{eq:3}) is consistent with the lower bound from \citet{Dermer}.
\Eqs{eq:1}{eq:3} give an example of fixing the dimensionless factors in
\Eq{argument}; the presence of $\alphaem$
indicates that this is a quantum effect, and the
ratio of relativistic degrees of freedom $g_{0}/g_\ast$ appears because
$n_\gamma$ dilutes as $T^3$ while the magnetic helicity decays with scale
factor $a$ as $a^{-3}$.

The results of our previous analysis allow us to determine the initial condition
for decaying helical turbulence.  To this end, we evaluate $v_\mu$ and
$v_\lambda$.
The above estimates of the maximal value of the chiral asymmetry give
\EQ
|\mu|\ll4\alphaem\frac{\kB T}{\hbar c}\approx
{1.5\times10^{14}}\,T_{100}\,\cm^{-1},
\EN
where $T_{100}$ is the temperature
in units of $100\GeV$ corresponding to $1.2\times10^{15}\K$.
For the magnetic resistivity we use Equation~(1.11) of \cite{Arnold00}:
\EQ
\eta={7.3\times 10^{-4}}\,{\hbar c^2\over\kB T}\approx
{4\times10^{-9}}T_{100}^{-1}\cm^2\s^{-1}.
\EN
Thus, $v_\mu={6 \times 10^5}\cm\s^{-1}$, so
the number of $e$-folds is ${\cal N}\equiv
v_\mu\mu/H\approx{5\times 10^{9}\, g_{100}^{-1/2} T_{100}^{-1}}\gg1$, where
$H^{-1}\approx5\times10^{-11}\,g_{100}^{-1/2}T_{100}^{-2}\,\s$ is the Hubble
time.

The chirality flipping rate is a complicated function of temperature below
$100\GeV$, with electromagnetic and weak processes as well as decays of
residual Higgs bosons contributing to it; see Appendix~D of \citet{BFR12}.
However, for a simple numerical estimate at $100\GeV$, we can extrapolate
the rate from the unbroken phase.
Using $\Gamma_{\rm\!f}/H\approx800\,T_{100}^{-1}$ \citep{CDEO92,JS97},
we find $\Gamma_{\rm f}/\eta\mu^2\approx(800/{\cal N})\,T_{100}^{-1}\approx
1.6\times10^{-7}g_{100}^{1/2}\ll1$. Although we underestimate the
flipping rate in this way, it remains negligible either way.

Next, we determine $\meanrho$ from the Friedmann equation:
$$\meanrho=\frac{\pi^2}{30}\,g_*\frac{(\kB T)^4}{\hbar^3c^5}\approx
7.6\times10^{26}g_{100}T_{100}^4\g\cm^{-3},$$ and we find
$$\lambda=3 \hbar c\,\left({8\alphaem\over\kB T}\right)^2\approx1.3\times10^{-17}\,T_{100}^{-2}\,\cm\erg^{-1}.$$
As a result, $v_\lambda\approx1.5\times10^{9}\cm\s^{-1} \gg v_\mu$ and
$v_\lambda\ll\cs\approx{}2\times10^{10}\cm\s^{-1}$, so we are in
regime~I where turbulence develops.
Finally, we estimate the length of the inertial range
of chiral magnetically driven turbulence from
\EQ
v_\mu/v_\lambda=\eta\,(\meanrho\lambda)^{1/2}\approx g_{100}^{1/2}/2400.
\EN
\Eq{klambda} with $\sqrt{C_\mu/C_\lambda}\approx4$ gives
$\mu/k_\lambda \approx 600 g_{100}^{-1/2}$.
So we expect a $k^{-2}$ spectrum covering almost three orders of magnitude.

\section{Conclusions}

We have shown that in chiral magnetohydrodynamics,
magnetic field evolution proceeds in
distinct stages:
(i) small-scale chiral dynamo instability;
(ii) first nonlinear stage when the Lorentz force drives
small-scale turbulence;
(iii) development of inverse energy transfer
with a $k^{-2}$ magnetic energy spectrum over the range $k_\lambda<k<\mu$;
and (iv) generation of large-scale magnetic field
by chiral magnetically driven turbulence,
decrease of the chemical potential, saturation,
and eventual decay.
This process acts as an intermediary to decaying hydromagnetic turbulence.

The application to the early universe results in a limit on magnetic
helicity, depending on the plasma particle content at high temperatures.
Larger values of $\bra{\BB^2}\,\xiM$ can only be envisaged if $\bra{\BB^2}$
and $\xiM$ are constrained separately, e.g., as a fraction of the radiation
energy of the universe and a fraction of the Hubble horizon \citep{KTBN13}.

\acknowledgments We thank Tanmay Vachaspati for useful discussions.  Support
through the NSF Astrophysics and Astronomy Grant Program (grant Nos.\ 1615100 and
1615940), the Research Council of Norway (FRINATEK grant No.\ 231444), the Georgian
NSF FR/264/6-350/14, and the European Research Council (grant number No.\ 694896) are
gratefully acknowledged.  We acknowledge the allocation of computing resources
provided by the Swedish National Allocations Committee at the Center for
Parallel Computers at the Royal Institute of Technology in Stockholm.  This
work utilized the Janus supercomputer, which is supported by the National
Science Foundation (award No.\ CNS-0821794), the University of Colorado
Boulder, the University of Colorado Denver, and the National Center for
Atmospheric Research. The Janus supercomputer is operated by the University of
Colorado Boulder.

%r e f

\end{document}